\documentclass{ws-ijmpa}
\usepackage[super,compress]{cite}
\usepackage{graphicx}
\UseRawInputEncoding 
\begin{document}
\markboth{G. L. Klimchitskaya, A. S. Korotkov, V. V. Loboda \& V. M. Mostepanenko}{Impact of Surface Roughness
on the Stability of Nanoelectromechanical Pressure
Sensors} 

%
\catchline{}{}{}{}{}
%

\title{\uppercase{Impact of surface roughness on the stability of nanoelectromechanical pressure
sensors in the Casimir regime}}

\author{\uppercase{G. L. Klimchitskaya,${}^{1,2}$ A. S. Korotkov,${}^2$
V. V. Loboda${}^2$} \lowercase{and} \hfill\\
\uppercase{V. M. Mostepanenko${}^{1,2}$}}
\address{${}^1$Central Astronomical Observatory at Pulkovo of the
Russian Academy of Sciences, Saint Petersburg,
196140, Russia\\
${}^2$Peter the Great Saint Petersburg
Polytechnic University, Saint Petersburg, 195251, Russia\\
g.klimchitskaya@gmail.com
}

\maketitle


\begin{abstract}
The stability of nanoelectromechanical pressure sensors working in the Casimir regime is
considered with account of surface roughness on both the sensor membrane and the
ground plate. The equilibrium positions of the sensor membrane are found from the balance
between the external measured, elastic, electric pressures, and the Casimir pressure computed
by means of the Lifshitz theory. It is shown that the stable equilibrium position of the sensor
membrane is nearly independent of the surface roughness, whereas its unstable equilibrium
position is shifted to larger membrane-plate separations. The use of these results for creating
pressure sensors with further shrinked dimensions is discussed.

\keywords{Surface roughness; pressure sensors; Casimir force.}
\end{abstract}

\ccode{PACS numbers: 12.20.Fv; 12.20.Ds; 68.35.Ct}


\newcommand{\kb}{{k_{\bot}}}
\newcommand{\xl}{{i\xi_l}}
\newcommand{\ve}{{\varepsilon}}
\newcommand{\okt}{{(\omega,k_t)}}

\section{Introduction}	

Recent trends are toward increased use of micro- and nanoelectromechanical sensors for measuring
different physical quantities in a scientific laboratory, industry, as well as in the everyday
life by means of capacitive, piezoelectric, optical and other techniques.\cite{1} A significant place
among all these microdevices is occupied by the pressure
sensors intended for work in both air and liquid environments.\cite{2,3}

The key role in the operation of microsensors is played by the electric and magnetic forces
described by classical electrodynamics. However, with shrinking the sensor sizes to below a
micrometer, the quantum effects caused by the electromagnetic fluctuations take effect as
was predicted by Richard Feynman 65 years ago\cite{4}. Specifically, at separations of a
few hundred nanometers the moving parts of a microsensor are under an impact
of the van der Waals (Casimir) force which exceeds in magnitude the typical electric forces
acting in this separation range.\cite{5,6,7} In the last few years, the Casimir force received
much experimental and theoretical attention\cite{7,8,9} and was used as a driving force in
micro- and nanoelecromechanical devices.\cite{10,11,12,13,14}

The role of the Casimir force in micro- and nanoelectromechanical pressure sensors was first
investigated in Ref.~\refcite{15}. It was shown that for some maximum value of the measured
pressure the sensor becomes unstable and collapses under an impact of the Casimir force.
For smaller measured pressures, the pressure sensor membrane has two equilibrium positions
one of which (at shorter separation of the sensor membrane from the ground plate) is
unstable and another one (at larger separation) is stable. The cases of a Si membrane and a Si
plate, as well as an Au-coated plate were considered.\cite{15} The pressure sensor with an
additional electric pressure between the sensor membrane and the ground plate was
demonstrated to have the properties like those in the absence of electric pressure.\cite{15}

In this paper, we examine the stability of the pressure sensors working in the Casimir regime
taking into account the role of surface roughness which was disregarded in Ref.~\refcite{15}.
It is shown that the position of a stable equilibrium of the sensor membrane calculated
with account of surface roughness remains essentially the same as for the smooth
surfaces. However, under an impact of surface roughness, the position of an unstable
equilibrium of the membrane is shifted to larger separations from the ground
plate. This shift is greater for larger roughness amplitudes.

\section{The Casimir Pressure with Account of Surface Roughness}

We consider the Casimir pressure in the configuration of the pressure sensor consisting of
the plane Si membrane of length $L=1000~\mu$m, width $D=200~\mu$m and
thickness $H=30~\mu$m, which is suspended at a height $h$ by the spring system above
either all-Si or an Au-coated Si plate.\cite{15} Let in the absence of the measured (external)
pressure, the weight of the membrane is balanced by the elastic force from the spring system
(the effective spring constant is $k$). Under an impact of the measured pressure $P$, the
membrane approaches the ground plate to so short separations that the Casimir force comes
into play.

The magnitude of the Casimir pressure between the smooth surfaces of a membrane and a plate
(both can be considered as infinitely thick) is presented by the Lifshitz theory\cite{16,17,18}
(see also Ref.~\refcite{7} for the more modern notations used here)
\begin{eqnarray}
&&
P_C(z,T)=\frac{k_BT}{\hbar}\sum_{l=0}^{\infty}{\vphantom{\sum}}^{\prime}\!\!
\int_{0}^{\infty}\!\!\!q_l\kb d\kb\sum_{\alpha}
\frac{r_{\alpha}^{(1)}(\xl,\kb)r_{\alpha}^{(2)}(\xl,\kb)\,e^{-2zq_l}}{1-
r_{\alpha}^{(1)}(\xl,\kb)r_{\alpha}^{(2)}(\xl,\kb)\,e^{-2zq_l}},
\label{eq1}
\end{eqnarray}
\noindent
where the first summation is over the Matrubara frequencies $\xi_l=2\pi k_BTl/\hbar$, $l=0,1,2$,...,
the prime on this sum divides the term with $l=0$ by 2, $k_B$ is the Boltzmann constant, and
$T$ is the temperature. The integration in (\ref{eq1}) is over the magnitude of
the wave vector projection on the plane of a membrane, $q_l=(k_{\bot}^2+\xi_l^2/c^2)^{1/2}$, and
$z$ is the separation distance between a membrane and a ground plate.

The second summation in (\ref{eq1}) is over the two polarizations of the electromagnetic field,
transverse magnetic ($\alpha$=TM) and transverse electric  ($\alpha$=TE) and $r_{\alpha}^{(p)}$
are the reflection coefficients on the membrane ($p=1$) and on the plate ($p=2$). They are
expressed as
\begin{eqnarray}
&&
r_{\rm TM}^{(p)}(\xl,\kb)=
\frac{\ve_l^{(p)}q_l-k_l^{(p)}}{\ve_l^{(p)}q_l+k_l^{(p)}}, \quad
r_{\rm TE}^{(p)}(\xl,\kb)=
\frac{q_l-k_l^{(p)}}{q_l+k_l^{(p)}},
\label{eq2}
\end{eqnarray}
\noindent
where $k_l^{(p)}=\left[k_{\bot}^2+\ve^{(p)}(\xl)\xi_l^2/{c^2}\right]^{1/2}$ and
$\ve^{(p)}(\xl)$ are the dielectric permittivities of a membrane and a plate materials calculated
at the pure imaginary Matsubara frequencies. If both a membrane and a plate materials are made
of Si, $\ve^{(1)}(\xl)=\ve^{(2)}(\xl)=\ve_{\rm Si}(\xl)$. If, however, the plate is coated with an
Au layer, which is usually of more than 100 nm thickness, it can be already considered as an
infinitely thick,\cite{7} so that $\ve^{(2)}(\xl)=\ve_{\rm Au}(\xl)$.

In actual conditions, the surfaces of both a membrane and a plate are rough. There are many
approaches to calculation of the Casimir force between rough surfaces.\cite{19,20,21,22}
The role of stochastic roughness with the amplitudes much smaller than the separation
distance $z$ and the correlation length much larger than $z$ can be taken into account
perturbatively. In this case, the magnitude of the Casimir pressure taking into account the impact
of surface roughness is given by\cite{23}
\begin{eqnarray}
P_C^R(z,T)=P_C(z,T)\left[1+10\frac{\delta_1^2+\delta_2^2}{z^2}+105\frac{(\delta_1^2+\delta_2^2)^2}
{z^4}\right],
\label{eq3}
\end{eqnarray}
\noindent
where $\delta_{1,2}$ are the r.m.s. roughness amplitudes on a membrane and on a plate, respectively,
counted from the zero roughness levels, whereas $P_C$ is given by Eq.~(\ref{eq1}).

\section{Equilibrium States of the Pressure Sensor with Rough Surfaces under an Impact of Elastic and
Casimir Forces}

In Ref.~\refcite{15}, the spring constant of the pressure sensor with smooth surfaces working in the
air environment was determined from the condition $kh=P_0LD$, where $P_0=3$~kPa and $h=20~\mu$m
is the equilibrium height where the membrane's weight is balanced by the elastic force. As a result,
$k=30~$N/m. In the absence of the Casimir force, the measured pressure $P=P_0$ would bring the
membrane supported by the spring system with this $k$ into the contact with the ground plate.

The equilibrium positions of the membrane under an impact of the measured pressure $P<P_0$ is reached
at some heights $z$ above the ground plate. These heights can be found from the condition
\begin{eqnarray}
f(z)\equiv\frac{k(h-z)}{LD}-P=P_C^R(z,T),
\label{eq4}
\end{eqnarray}
\noindent
which means that the elastic pressure is balanced by the sum of the measured and Casimir pressures.

First we consider the case of a Si membrane and a Si plate (here and below at $T=300$~K). The quantity
$\ve_{\rm Si}(\xl)$ is found by the Kramers-Kronig relations using the optical data for
Im$\ve_{\rm Si}(\omega)$ taken
for the high-resistivity dielectric samples (note that at separations below 200~nm considered here the free
charge carriers do not influence the Casimir pressure if their concentration is below the critical
one).\cite{25,26,27}

Computations were performed by (\ref{eq1}), (\ref{eq3}), and (\ref{eq4}). The function $f(z)$ was
calculated with the measured pressure $P=2.979$~kPa slightly smaller than that leading to the sensor collapse.
The Casimir pressure $P_C^R$ was computed with the r.m.s. roughness amplitude $\delta_{\rm Si}=0.1$~nm,
as measured in the experiment.\cite{25} The equilibrium positions of the membrane are given by the two
solutions of Eq.~(\ref{eq4}). The first solution $z_{\rm eq}^{\rm unst}\approx75.5$~nm
is the position of an unstable equilibrium, whereas the second one, $z_{\rm eq}^{\rm st}\approx132$~nm, is
the stable one. These solutions are the same as obtained earlier\cite{15} for smooth surfaces. Thus, in this
case, roughness does not make any impact on the equilibrium positions of a membrane due to extremely smooth
character of Si surfaces used in Ref.~\refcite{25}.

Next we consider the case of a Si membrane above an Au-coated plate. The permittivity $\ve_{\rm Au}(\xl)$
is found using the optical data\cite{24} extrapolated to zero frequency.
At the separations considered, the type of extrapolation does not influence the obtained results.\cite{7}
The r.m.s. roughness amplitude $\delta_{\rm Au}=2$~nm has been used in computations.\cite{28} Two
obtained equilibrium positions of the sensor membrane, $z_{\rm eq}^{\rm unst}\approx83$~nm and
$z_{\rm eq}^{\rm st}\approx129.5$~nm, again coincide with those computed earlier for
the smooth Si and Au surfaces.\cite{15}

\begin{figure}[b]
\vspace*{-9.5cm}
\hspace*{-2.5mm}
\centerline{\includegraphics[width=15.5cm]{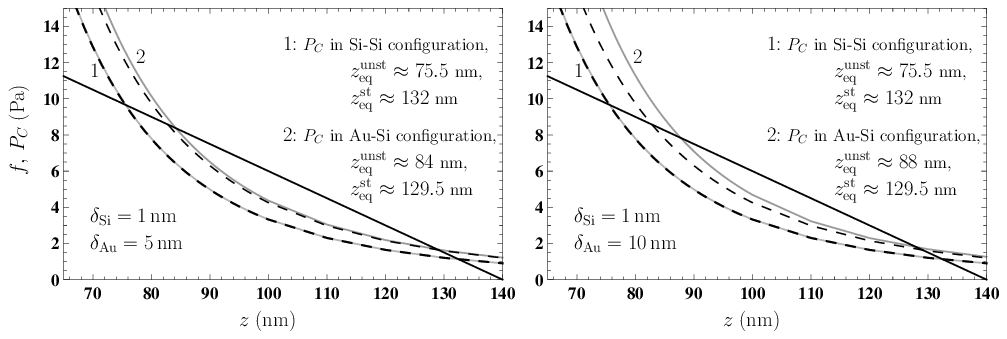}}
\vspace*{-8.6cm}
\caption{Unstable and stable equilibrium positions of the rough Si membrane above the rough surface of either
Si or Au-coated Si plate in the presence of the Casimir pressure are shown by intersections of the solid lines
1 and 2, respectively, plotted for $\delta_{\rm Si}=1$~nm, $\delta_{\rm Au}=5$~nm (left) and
$\delta_{\rm Si}=1$~nm, $\delta_{\rm Au}=10$~nm (right) with the straight line. The dashed lines 1 and 2
are plotted for the smooth surfaces.
 \label{f1}}
\end{figure}
In experiments on measuring the Casimir force,\cite{25,28} the special measures have been undertaken
in order to decrease the surface roughness. Because of this, it is important to determine the impact of surface
roughness on the stability of a pressure sensor with greater roughness amplitudes. In Fig. 1, the
computational results for the function $f(z)$ are shown by the straight lines, but the solid lines
1 and 2 are computed for $\delta_{\rm Si}=1$~nm, $\delta_{\rm Au}=5$~nm (left) and
$\delta_{\rm Si}=1$~nm, $\delta_{\rm Au}=10$~nm (right). These amplitudes exceed the ones
used above by the factor of 10 for Si and by the factors of 2.5 in Fig. 1 (left) and 5 in Fig. 1 (right) for Au.
The dashed lines 1 and 2 are plotted with no account of surface roughness.

As is seen in Fig. 1, for a Si membrane above a Si ground plate even by the order of magnitude larger
roughness does not influence the equilibrium positions. This is not the case, however, for a Si membrane
above an Au-coated plate. From Fig. 1 (left) it is seen that although the position of a stable equilibrium
remains unchanged, $z_{\rm eq}^{\rm st}\approx129.5$~nm, the unstable one
is shifted for 1~nm to a larger separation, $z_{\rm eq}^{\rm unst}\approx84$~nm (see the solid lne 2
which deviates from the dashed one).

In Fig. 1 (right), where $\delta_{\rm Au}=10$~nm, the position of an unstable equilibrium for a Si
membrane above an Au plate is shifted for 4~nm to larger separations as compared to the
case of smooth surfaces (see the solid and dashed lines labeled 2). To conclude, the surface
roughness on Au surfaces influences the position of an unstable equilibrium of the sensor
membrane by shifting it to larger separations.

\section{Equilibrium States of the Pressure Sensor with Rough Surfaces under an Impact of Electric,
Elastic and Casimir Forces}

The application of some potential difference $U_0$ between a sensor membrane and a ground plate provides
a way of determining membrane displacements by means of the capacitive measurements. The magnitude
of the electric force between a membrane and a ground plate with account of surface roughness is given by
\begin{eqnarray}
P_{\rm el}^R(z)=\frac{\epsilon_0U_0^2}{2z^2}
\left[1+3\frac{\delta_1^2+\delta_2^2}{z^2}+5\frac{(\delta_1^2+\delta_2^2)^2}{z^4}\right],
\label{eq5}
\end{eqnarray}
\noindent
where $\epsilon_0$ is the permittivity of free space. As a result, instead of (\ref{eq4}), the equilibrium height
$z$ of the membrane above the ground plate is found from the condition
\begin{eqnarray}
f(z)\equiv\frac{k(h-z)}{LD}-P=P_C^R(z,T)+P_{\rm el}^R(z)\equiv P_{\rm tot}^R(z).
\label{eq6}
\end{eqnarray}

As in the previous section, $k=30$~N/m. Computations were performed with the measured pressure of
$P=2.973$~kPa and the experimentally measured\cite{25,28} r.m.s. roughness amplitudes
$\delta_{\rm Si}=0.1$~nm and $\delta_{\rm Au}=2$~nm. The obtained equilibrium positions of the
sensor membrane  $z_{\rm eq}^{\rm unst}\approx84$~nm, and  $z_{\rm eq}^{\rm st}\approx164.5$~nm
coincide with those obtained in Ref.~\refcite{15} with no account of surface roughness.

Now we consider the same pressure sensor with larger r.m.s. roughness amplitudes.
The measured pressure is again $P=2.973$~kPa. The computational
results for the total pressure $P_{\rm tot}^R$ and the function $f$ are presented in Fig. 2 by the
solid curved and straight lines, respectively, plotted for  $\delta_{\rm Si}=1$~nm,
$\delta_{\rm Au}=5$~nm (left) and  $\delta_{\rm Si}=1$~nm, $\delta_{\rm Au}=10$~nm (right).
The dashed lines show $P_{\rm tot}$ computed with no account of surface roughness.

\begin{figure}[t]
\vspace*{-9.2cm}
\hspace*{-2.5mm}
\centerline{\includegraphics[width=15.5cm]{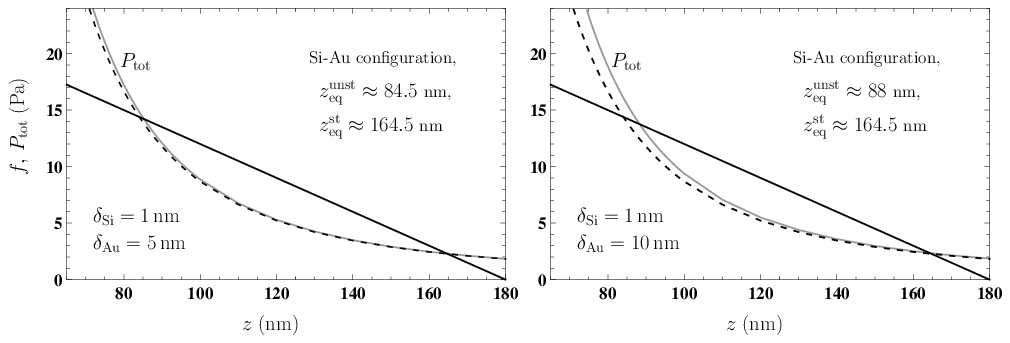}}
\vspace*{-8.6cm}
\caption{Unstable and stable equilibrium positions of the rough Si membrane above the rough surface of
Au-coated Si plate in the presence of the Casimir and electric pressures are shown by intersections of the solid
curved line plotted for $\delta_{\rm Si}=1$~nm, $\delta_{\rm Au}=5$~nm (left) and
$\delta_{\rm Si}=1$~nm, $\delta_{\rm Au}=10$~nm (right) with the straight line. The dashed lines
are plotted for the smooth surfaces.
 \label{f2}}
\end{figure}

Now we consider the same pressure sensor with larger r.m.s. roughness amplitudes.
The measured pressure is again $P=2.973$~kPa. The computational
results for the total pressure $P_{\rm tot}^R$ and the function $f$ are presented in Fig. 2 by the
solid curved and straight lines, respectively, plotted for  $\delta_{\rm Si}=1$~nm,
$\delta_{\rm Au}=5$~nm (left) and  $\delta_{\rm Si}=1$~nm, $\delta_{\rm Au}=10$~nm (right).
The dashed lines show $P_{\rm tot}$ computed with no account of surface roughness.

As is seen in Fig. 2 (left), the position of a stable equilibrium remains unchanged whereas the position
of an unstable one is shifted for only 0.5~nm. From Fig. 2 (right) one again finds that the position of a
stable equilibrium is unchanged but the position of an unstable equilibrium is shifted for 4~nm to larger
separations. Thus, in the presence of electric force, the account of surface roughness leads to
qualitatively the same results regarding the equilibrium positions of sensor membrane as were obtained
in Sec. 3 with no electric force.

\section{Conclusions and Discussion}

In the foregoing, we considered the nanoelectromechanical pressure sensor working in the Casimir regime
with account of surface roughness on the surfaces of a sensor membrane and a ground plate. It was shown
that the position of a stable equilibrium of the membrane is not influenced by the presence of
surface roughness. However, an account of roughness shifts the position of an unstable equilibrium
of the membrane to larger separations from the plate. This shift is negligibly small for rather smooth
surfaces used in precise experiments on measuring the Casimir force performed in high vacuum but
becomes more sizable with larger roughness amplitudes.

One should take into account that the micro- and nanoelectromechanical pressure sensors of the next
generations with further shrinked dimensions should work not in a high vacuum of precise experiments and,
preferably, should not use the expressly designed highly smooth surfaces. Because of this, the obtained
results regarding the impact of surface roughness on the state of sensor equilibrium may be helpful
in their development and improvement.

\section*{Acknowledgments}

This work was supported by the
State Assignment for basic research (project FSEG--2023--0016).

\end{document}